\documentclass[10pt, oneside, a4paper, twocolumn]{article}

 \usepackage{rotating}
\usepackage[cjk]{kotex}
\usepackage{setspace}
\usepackage[a4paper,left=20mm,right=20mm,top=30mm,bottom=30mm]{geometry}
\usepackage{tikz}
\usepackage[letterspace=200]{microtype}
\usepackage{graphicx}
\usepackage{amssymb}
\usepackage{endnotes}
\usepackage{amsmath}
\usepackage[normalem]{ulem}
\useunder{\uline}{\ul}{}
\usepackage{booktabs}
\graphicspath{ {fig/} }
\usepackage{tabularx}
\usepackage{array}
\newcolumntype{Y}{>{\centering\arraybackslash}X}
\usepackage{fancyhdr}
 
\begin{document}

\onecolumn
\setstretch{1.2}

\title{Do We Price Happiness? Evidence from Korean Stock Market}

\author{HyeonJun Kim$^{*}$\medskip\\{\normalsize School of Finance, Soongsil University}
\medskip\\{\normalsize$^*$inertia164214@soongsil.ac.kr or hyeonjunacademic@gmail.com}}


\maketitle

\begin{abstract}

This study explores the potential of internet search volume data, specifically Google Trends, as an indicator for cross-sectional stock returns. Unlike previous studies, our research specifically investigates the search volume of the topic 'happiness' and its impact on stock returns in the aspect of risk pricing rather than as sentiment measurement. Empirical results indicate that this 'happiness' search exposure (HSE) can explain future returns, particularly for big and value firms. This suggests that HSE might be a reflection of a firm's ability to produce goods or services that meet societal utility needs. Our findings have significant implications for institutional investors seeking to leverage HSE-based strategies for outperformance. Additionally, our research suggests that, when selected judiciously, some search topics on Google Trends can be related to risks that impact stock prices.
\\\\

\textbf{keywords:} {[Asset Pricing], [Sentiment], [Google Trend]}


\end{abstract}

\setstretch{1.5}

\section{Introduction}

Market sentiment is shown to be regraded as the predictor for cross-sectional stock return, even when adjusted by size, value, and momentum effect. \cite{baker2006investor} This initial way of measuring market sentiment by Baker et al. (2006) was based on market-based measures, while others used suvey-based indicies. \cite{brown2005investor} Later an alternative is suggested that use internet search volume such as Google Trends \cite{da2015sum}. This method is quite appealing for financial economists because of its transparency of models built upon it, while also obtaining high frequency data. There was some concerns about the quality of Google Trend data \cite{cebrian2023google}, however it is certain that Google Trend data could be used as "early warning signs of stock market moves" \cite{preis2013quantifying}. Here the natural question arises; could google search queries detect fundamentals and risk associated with it?

Usually, search volume or related metrics (like return sensitivity) is far from fundamental information, and rather a sentiment proxy. This is because search volume, by its nature, is about collective social attention. However, it is shown that exposure, or factor loading to this sentiment proxy can be used for variety of asset pricing application. There are attempts to use this exposure by Chen, Kumar and Zhang (2020), but the study focused on relationship between the factor loading of a search volume and abnormal return\cite{chen2020dynamic}, which is within the area of mispricing or irrational asset pricing.
In this paper we suggest the possibility that internet search volume data could be used for rational asset pricing by extracting the common risk across assets. Empirical analysis show that stock return's exposure to the search volume of topic 'happiness'(or HSE) have explanatory power to future return even when BE/ME and size is controlled. By subsampling the firms by size and BE/ME, the predictive power of HSE is higher in big and value firms, supporting the theory that HSE is a proxy for firm's capacity to generate goods or service that satisfies society's utility, or happiness.

\section{What is Fundamental about Sentiment?}

As any search query could be chosen to be used as demand pressure, at least it will be interesting, if not useful, to measure the demand pressure regarding abstract subject. Fundamentally, the search engines are for searching information or solution of some needs. For example, query volume for 'jobs' will have thorough economic explanation. Whereas search queries like "Happiness" do not have clear economic meaning. Usually users could not use the search engine to relief its demand about abstract concepts. Rather it should be interpreted as general sentiment about those concepts. Then we can intuitively interpret the high or low sentiment for these concepts in its own context of society, and check whether or not the sentiment proxies to certain economic activities. High sentiment for "Happiness", for instance, means there is overall lack of happiness in the society. Therefore, search volume for "Happiness" could be regarded as demand for happiness.

The time series format of this search volume intensity enables to measure quantity of exposure for these sentiments using Ross (1976)'s method. \cite{ross1976arbitrage} The discussion that should be address is whether or not this this exposure (or Happiness Sentiment Exposure, HSE) could be considered as risk. This question could be solved by empirical evidence that will be presented later in this paper, but that do not give confidence in theoretical level. Here we focus on the fact that as customer utility is generated by happiness, the overall demand of happiness in the society influence the customer demand of the product that the some company serves, thus influence the fundamental of the company. Therefore, we hypothesized that Happiness Sentiment Exposure measured as the factor loading for search volume intensity for "Happiness(행복)" is somewhat theoretically adequate to be regarded as risk variables.
We used a exposure to the search volume intensity of the subject 'happiness' that Google Trend provide to estimate the risk proxy for each stock. This risk proxy will quantify the tendency of the firm to be more related to society's utility supply processing. 

\section{Empirical Analysis}

We use the cross-sectional analysis method of Fama and French (1992) \cite{fama1992cross} for identifying cross-sectional excess return pattern related to HSE. Although later studies did identified few more market anomalies like momentum, profitability and investment, 

\subsection{Data Preparation}

Daily return including dividends, daily market cap of common stocks, equity, preferred stock data was collected from DataGuide. Return data is changed from daily to monthly frequency, and BE/ME at year $t$ was calculated by dividing equity minus preferred stock last trading day market cap of December of year $t$

The model for extracting the exposure $\beta_{SVI}$ is presented in equation \ref{exp}. 
\begin{equation}\label{exp}
    R_{t}^{(i)} = \alpha + \beta_{SVI}\Delta SVI_ {t} + \epsilon_t
\end{equation}
Where $R_{t}^{(i)}$ is monthly return of a stock $i$ at time $t$ and $\Delta SVI_t$ is log difference between monthly search volume intensity of query related to "happiness" between time $t$ and $t-1$. Also, although we mentioned the model of Ross (1976), equation \ref{exp} does not include market proxy for a few reasons. First, as we assumed that Happiness Sentiment Exposure is systematical but noisy, we had a concern that Happiness Sentiment Exposure will be misestimated due to market proxy, which have stronger relationship with the returns and is noise-free independent variable. Second, if the market proxy is added to the analysis, then the factor loading for Happiness Sentiment will measure the relationship with Happiness Sentiment with abnormal returns, which is not this paper's topic. 
The returns are winsorized by 99percentile and 1percentile threshold values, and estimation data range is 72 months, but data length down to 24 month was also accepted.

For portfolio analysis and Fama-Macbeth regression, the return was symmetrically trimmed by one observation each month to exclude extreme returns.

\subsection{Preliminary}

Here we will address the local characteristics of Korean stock market. While it is not thoroughly researched, the two main stock market in Korea, KOSPI market and KOSDAQ seem less likely to be stochastically segmented or equally valuated. Unlike NYSE and NASDAQ, KOSDAQ stocks tends to transfer to KOSPI if possible, and index value of KOSDAQ, which started at 1000 in year 1997, still didn't fully recovered its initial value, while KOSPI market grow more than 200\%. Considering these facts, it will be wise to choose one of the two markets for researching asset pricing that will avoid inconsistent results. We choose KOSPI, which has more larger market capitalization.

\subsection{Informal Tests}

Here we first report portfolio analysis results for presenting non-parametric results and finding some discussion points. The result is shown in Table \ref{table:1}

\begin{table}[h!]
\label{table:1}

\centering
\begin{tabularx}{\textwidth}{X|YYYY}
\hline
\multicolumn{1}{l}{\textbf{}}              & \textbf{HSE}             & \textbf{BE/ME(+)}           & \textbf{Size}              & \textbf{Excess Return}                                       \\ \hline
\textbf{Unhappy}                           & -0.19                    & 1.25                     & 847.3                      & \begin{tabular}[c]{@{}r@{}}-1.04\\      (-1.81)\end{tabular} \\
\textbf{2}                                 & -0.11                    & 1.76                     & 1094.46                    & \begin{tabular}[c]{@{}r@{}}0.25\\      (0.54)\end{tabular}   \\
\textbf{3}                                 & -0.07                    & 1.81                     & 1464.26                    & \begin{tabular}[c]{@{}r@{}}0.26\\      (0.62)\end{tabular}   \\
\textbf{4}                                 & -0.05                    & 1.94                     & 1931.27                    & \begin{tabular}[c]{@{}r@{}}0.47\\      (1.09)\end{tabular}   \\
\textbf{5}                                 & -0.03                    & 1.76                     & 1839.11                    & \begin{tabular}[c]{@{}r@{}}0.37\\      (0.78)\end{tabular}   \\
\textbf{6}                                 & -0.01                    & 1.75                     & 1362.77                    & \begin{tabular}[c]{@{}r@{}}-0.10\\      (-0.23)\end{tabular} \\
\textbf{7}                                 & 0                        & 1.71                     & 2167.74                    & \begin{tabular}[c]{@{}r@{}}0.65\\      (1.54)\end{tabular}   \\
\textbf{8}                                 & 0.03                     & 1.69                     & 1576.52                    & \begin{tabular}[c]{@{}r@{}}0.27\\      (0.63)\end{tabular}   \\
\textbf{9}                                 & 0.05                     & 1.47                     & 1262.08                    & \begin{tabular}[c]{@{}r@{}}0.89\\      (2.09)\end{tabular}   \\
\textbf{Happy}                             & 0.12                     & 1.37                     & 1446.29                    & \begin{tabular}[c]{@{}r@{}}0.86\\      (1.84)\end{tabular}   \\
\textbf{Happy-Unhappy} & 0.31 & 0.12 & 598.99 & \begin{tabular}[c]{@{}r@{}}1.90\\      (3.65) 
\end{tabular}  \\
\hline
\end{tabularx}
\caption{Univariate sort result of Happiness Sentiment Exposure(HSE). The portfolio is formed by sorting the stocks with cross-sectional decile of HSE for each year. we report time-series average for average BE/ME and market capitalization(market cap) of each stock in the portfolio, in which market cap is reported in 1 billion Korean won. Portfolio return is calculated by value-weighting the returns with market cap and subtracting with risk-free rate. We used monthly investment return of 1-year Korean Government treasury bill for risk-free rate. The portfolio is formed in June of year $t$ using the HSE data from July of year $t-4$ to June of year $t$(with at least 24 observation) and reblanced that July of year $t+1$.}
\end{table}

Here we can see some expected, and also interesting results. First, as expected we see a somewhat monotonic increase in average excess return when the HSE value increases. T-values is not statistically significant (the highest decile being 1.67 standard deviation away from 0), but we can argue that this is because of noisy return, not because of the true average return value. After 8th decile, annualized expected return is over 9.3\%, and the annualized excess return difference between 10th decile(Happy) and 1st decile(Unhappy) is over 17\%. We did not report the equally weighted portfolio excess return, but we want to note that for the equally weighted portfolio, the excess return for high decile portfolio is robust and statistically significant, but the monotonic increase of return in higher decile became less apparent. We assume that this is due to size effect and value effect that is observed in the Korean stock market, i.e. the high deciles have low BE/ME and low decile is small in size thus the monotonic pattern are flatten. \\
We also can observe unusual results for average stock market cap for each decile, which is non-linear. Actually the middle deciles (around 4-7th) has the highest BE/ME and market cap. This leads to interesting interpretation; low HSE stocks and high HSE stocks are both small cap and low value firms, while middle HSE stocks are big cap and high value firms. According to previous studies that argues that KOSPI exchange stocks have little or reversed size effect\cite{eom2014reexamination}, the middle HSE portfolio should be the most high performing portfolios. However the sorting result shows that HSE does have its own market anomaly phenomenon regardless of pre-existing anomalies. 
Because of the previously reported non-linear relationship between BE/ME or size and HSE, We also report the 5 by 5 size-HSE and BE/ME-HSE double sort result to show HSE's predictive power of excess returns when size or BE/ME is controlled. Specific method is same as beta-size sort of Fama\&French(1992) \cite{fama1992cross}. Stocks in each size or BE/ME decile portfolio is sorted again according to HSE's decile breakpoints. Here, we found that in big firm or high value firms subsample, the result more strongly supports our hypothesis that HSE is a risk proxy, or at least a persisting anomaly unrelated to size or BE/ME. 

\begin{table}
\centering
\begin{tabularx}{\textwidth}{XYYYYYY}
\hline
\multicolumn{1}{l}{\textbf{}} & \textbf{Unhappy}                                             & \textbf{2}                                                 & \textbf{3}                                                 & \textbf{4}                                                 & \textbf{Happy}                                             & \textbf{Happy-Unhappy}                                \\ \hline
\textbf{Small}                & \begin{tabular}[c]{@{}r@{}}1.74\\      (3.24)\end{tabular}   & \begin{tabular}[c]{@{}r@{}}2.21\\      (4.22)\end{tabular} & \begin{tabular}[c]{@{}r@{}}2.15\\      (4.37)\end{tabular} & \begin{tabular}[c]{@{}r@{}}1.69\\      (3.12)\end{tabular} & \begin{tabular}[c]{@{}r@{}}1.84\\      (3.51)\end{tabular} & \begin{tabular}[c]{@{}r@{}}0.10\\ (0.24)\end{tabular} \\
\textbf{2}                    & \begin{tabular}[c]{@{}r@{}}1.22\\      (2.46)\end{tabular}   & \begin{tabular}[c]{@{}r@{}}1.33\\      (2.91)\end{tabular} & \begin{tabular}[c]{@{}r@{}}0.97\\      (2.11)\end{tabular} & \begin{tabular}[c]{@{}r@{}}1.45\\      (3.12)\end{tabular} & \begin{tabular}[c]{@{}r@{}}1.15\\      (2.24)\end{tabular} & \begin{tabular}[c]{@{}r@{}}-0.07\\ (-0.24)\end{tabular} \\
\textbf{3}                    & \begin{tabular}[c]{@{}r@{}}0.71\\      (1.41)\end{tabular}   & \begin{tabular}[c]{@{}r@{}}0.63\\      (1.37)\end{tabular} & \begin{tabular}[c]{@{}r@{}}0.88\\      (1.88)\end{tabular} & \begin{tabular}[c]{@{}r@{}}1.12\\      (2.34)\end{tabular} & \begin{tabular}[c]{@{}r@{}}0.90\\      (1.82)\end{tabular} & \begin{tabular}[c]{@{}r@{}}0.20\\ (0.73)\end{tabular} \\
\textbf{4}                    & \begin{tabular}[c]{@{}r@{}}-0.03\\      (-0.05)\end{tabular} & \begin{tabular}[c]{@{}r@{}}0.53\\      (1.12)\end{tabular} & \begin{tabular}[c]{@{}r@{}}0.86\\      (2.01)\end{tabular} & \begin{tabular}[c]{@{}r@{}}0.73\\      (1.81)\end{tabular} & \begin{tabular}[c]{@{}r@{}}0.93\\      (1.94)\end{tabular} & \begin{tabular}[c]{@{}r@{}}0.96\\ (3.12)\end{tabular} \\
\textbf{Big}                  & \begin{tabular}[c]{@{}r@{}}0.09\\      (0.18)\end{tabular}   & \begin{tabular}[c]{@{}r@{}}0.45\\      (1.01)\end{tabular} & \begin{tabular}[c]{@{}r@{}}0.91\\      (2.28)\end{tabular} & \begin{tabular}[c]{@{}r@{}}0.64\\      (1.59)\end{tabular} & \begin{tabular}[c]{@{}r@{}}0.85\\      (2.07)\end{tabular} & \begin{tabular}[c]{@{}r@{}}0.76\\ (2.23)\end{tabular} \\  \hline
\end{tabularx}
\caption{Sort result of stocks in KOSPI with quintiles of Size and HSE variable. For each size quintile, HSE quintile breakpoint is used to form HSE quintile portfolios inside the same size quintile. Size and HSE are observed in June of year $t$. The portfolio is formed and rebalanced on the last day of June of year $t$.}
\label{table:2}
\end{table}

\begin{table}[h!]
\centering
\begin{tabularx}{\textwidth}{XYYYYYY}
\hline

\textbf{}     & \textbf{Unhappy}                       & \textbf{2}                             & \textbf{3}                           & \textbf{4}                           & \textbf{Happy}                       & \textbf{Happy-Unhappy}                 \\ \hline
\textbf{Low}  & \begin{tabular}[c]{@{}c@{}}0.16\\      (0.31)\end{tabular} & \begin{tabular}[c]{@{}c@{}}0.54\\      (1.23)\end{tabular} & \begin{tabular}[c]{@{}c@{}}0.34\\      (0.78)\end{tabular} & \begin{tabular}[c]{@{}c@{}}0.57\\      (1.26)\end{tabular} & \begin{tabular}[c]{@{}c@{}}0.57\\      (1.22)\end{tabular} & \begin{tabular}[c]{@{}c@{}}0.41\\      (1.18)\end{tabular} \\
\textbf{2}    & \begin{tabular}[c]{@{}c@{}}0.69\\      (1.48)\end{tabular} & \begin{tabular}[c]{@{}c@{}}0.79\\      (1.77)\end{tabular} & \begin{tabular}[c]{@{}c@{}}0.95\\      (2.07)\end{tabular} & \begin{tabular}[c]{@{}c@{}}0.88\\      (1.98)\end{tabular} & \begin{tabular}[c]{@{}c@{}}0.73\\      (1.59)\end{tabular} & \begin{tabular}[c]{@{}c@{}}0.03\\      (0.10)\end{tabular} \\
\textbf{3}    & \begin{tabular}[c]{@{}c@{}}1.11\\      (2.30)\end{tabular} & \begin{tabular}[c]{@{}c@{}}1.24\\      (2.80)\end{tabular} & \begin{tabular}[c]{@{}c@{}}1.17\\      (2.80)\end{tabular} & \begin{tabular}[c]{@{}c@{}}0.91\\      (1.89)\end{tabular} & \begin{tabular}[c]{@{}c@{}}1.44\\      (2.89)\end{tabular} & \begin{tabular}[c]{@{}c@{}}0.32\\      (1.09)\end{tabular} \\
\textbf{4}    & \begin{tabular}[c]{@{}c@{}}1.09\\      (2.28)\end{tabular} & \begin{tabular}[c]{@{}c@{}}1.28\\      (2.71)\end{tabular} & \begin{tabular}[c]{@{}c@{}}1.48\\      (3.08)\end{tabular} & \begin{tabular}[c]{@{}c@{}}1.52\\      (3.25)\end{tabular} & \begin{tabular}[c]{@{}c@{}}1.44\\      (2.93)\end{tabular} & \begin{tabular}[c]{@{}c@{}}0.36\\      (1.25)\end{tabular} \\
\textbf{High} & \begin{tabular}[c]{@{}c@{}}1.20\\      (2.51)\end{tabular} & \begin{tabular}[c]{@{}c@{}}1.39\\      (3.05)\end{tabular} & \begin{tabular}[c]{@{}c@{}}1.36\\      (3.05)\end{tabular} & \begin{tabular}[c]{@{}c@{}}1.84\\      (3.80)\end{tabular} & \begin{tabular}[c]{@{}c@{}}1.66\\      (3.28)\end{tabular} & \begin{tabular}[c]{@{}c@{}}0.46\\      (1.51)\end{tabular} \\  \hline
\end{tabularx}
\caption{Sort result of stocks in KOSPI with quintiles of BE/ME and HSE variable.  For each BE/ME quintile, HSE quintile breakpoint is used to form HSE quintile portfolios inside the same BE/ME quintile. Size and HSE are observed at June of year $t$,. The portfolio is formed at June of year $t$ and rebalanced at June of year $t+1$.}
\label{table:3}
\end{table}

\subsection{Regression Analysis}

Table \ref{tab4} show the Fama-Macbeth Regression result for several variables. The regression result will show some insights about the HSE's predictability, and also the conditional predictability when the other variables are controlled.
\begin{table}[h!]
\centering
\begin{tabular}{lllll}

\hline
\textbf{a} & \textbf{s} & \textbf{h} & \textbf{$h_{dummy}$} & \textbf{hs} \\ \hline
3.69       & -0.22      &            &                   &             \\
3.49       & -2.79      &            &                   &             \\
           &            &            &                   &             \\
0.77       &            & 0.70       & -2.22             &             \\
1.84       &            & 6.75       & -2.43             &             \\
           &            &            &                   &             \\
0.88       &            &            &                   & 2.20        \\
2.13       &            &            &                   & 2.01        \\
           &            &            &                   &             \\
2.83       & -0.17      & 0.60       & -2.47             &             \\
0.92       & -2.18      & 5.34       & -2.77             &             \\
           &            &            &                   &             \\
4.19       & -0.26      &            &                   & 2.08        \\
4.05       & -3.94      &            &                   & 1.97        \\
           &            &            &                   &             \\
0.77       &            & 0.45       & -2.61             & 2.02        \\
1.90       &            & 3.75       & -1.92             & 1.82        \\
           &            &            &                   &             \\
3.52       & -0.21      & 0.32       & -2.83             & 2.12        \\
3.41       & -3.17      & 2.68       & -2.07             & 2.00        \\
           &            &            &                   &            \\
\hline
\end{tabular}
\caption{On each month, we conducted the following regression. \\ $R_i-RF=a + s  \cdot log(ME_i) + h \cdot log(BE/ME)_i^{+} + h_{dummy} \cdot BE_{Dummy, i} + hs \cdot HSE_i + \epsilon_i $.\\ $log(BE/ME)^{+}_i$ is BE/ME value that fills BE/ME with BE less than 0. $BE_{Dummy, i}$ is 1 when BE is less than 0, else 1. Then we calculated time-series average and its t-value of coefficients and intercept. Returns are observed at July, year $t$ to June, year $t+1$, and BE/ME are observed at December, year $t-1$ and HSE, ME is observed at June, year $t$.}
\label{tab4}
\end{table}

Here we can see that HSE does explain some cross-sectional return in univariate condition. The t-value of HSE coefficient is statistically significant by a margin, but considering that HSE has estimation error, this result can be seen as robust. bivariate regression of HSE and $log(BE/ME)$ (more precisely $log(BE/ME)^{+}_i$, HSE, and $BE_{Dummy, i}$) show the decrease in HSE's explanatory power by a lot. \\
Interesting observation is shown when we consider the relationship between HSE and $log(ME)$. When comparing 4th and 7th regression, the significance of ME change dramatically, from 2.18 standard deviation away from 0 to 3.17 standard deviation away from 0, with increase in risk premium as well. This is also shown in 1st amd 5th regression. This result could be understood by HSE-size sort result that shown previously. If the HSE can explain big firm's abnormal high performance in Korean stock market \cite{eom2014reexamination}, the size effect become more apparent. 
Whereas, HSE seems to not sufficiently explain high value firms' performance, though the risk premium for BE/ME lessen.
The regression that includes all variables show that HSE does have explanatory power when size and BE/ME is controlled.
\begin{table}[h!]
\centering
\begin{tabular}{llllll}
\hline
\textbf{Sub Sample}    & \textbf{a} & \textbf{s} & \textbf{h} & \textbf{$h_{dummy}$} & \textbf{hs} \\ \hline
Big   & 11.37   & -0.74   & 0.10     & -0.94          & 3.58    \\
                       & (8.08)   & (-8.13)   & (0.71)  & (-0.60)          & (2.42)    \\
                       &            &            &            &                   &             \\
Small & 23.40   & -2.06   & 0.69   & -4.23          & 2.08     \\
                       & (13.06)   & (-13.45)   & (4.96)   & (-4.21)          & (1.46)    \\
                       &            &            &            &                   &             \\ \hline
\end{tabular}
\caption{Big and Small subsample of firms are determined by yearly median of size(log value of market cap) value. If the firm's size at year $t$ is bigger than the size median at time $t$, is classified as big firm, and vise versa. On each month, we conducted the following regression. \\ $R_i-RF=a + s  \cdot log(ME_i) + h \cdot log(BE/ME)_i^{+} + h_{dummy} \cdot BE_{Dummy, i} + hs \cdot HSE_i + \epsilon_i $.\\ $log(BE/ME)^{+}_i$ is BE/ME value that fills BE/ME with BE less than 0. $BE_{Dummy, i}$ is 1 when BE is less than 0, else 1. Then we calculated time-series average and its t-value of coefficients and intercept. Returns are observed at July, year $t$ to June, year $t+1$, and BE/ME are observed at December, year $t-1$ and HSE, ME is observed at June, year $t$.}
\label{sb}
\end{table}

\begin{table}[h!]
\centering
\begin{tabular}{llllll}
\hline
\textbf{Sub Sample}     & \textbf{a} & \textbf{s} & \textbf{h} & \textbf{$h_{dummy}$} & \textbf{hs} \\ \hline
Value  & 5.58       & -0.37      & -0.06      & 0.00                 & 3.94        \\
                        & 4.68       & -4.35      & -0.33      & 1.48              & 2.58        \\
                        &            &            &            &                   &             \\
Growth & 2.24       & -0.11      & 0.40       & 0.00              & 0.96        \\
                        & 1.97       & -1.49      & 2.43       & -0.89             & 0.79        \\
                        &            &            &            &                   &             \\ \hline
\end{tabular}
\caption{Value and Growth subsample of firms are determined by yearly median of BE/ME value (including negative BE/ME). If the firm's BE/ME at year $t$ is bigger than the BE/ME median at time $t$, is classified as big firm, and vise versa. On each month, we conducted the following regression. \\ $R_i-RF=a + s  \cdot log(ME_i) + h \cdot log(BE/ME)_i^{+} + h_{dummy} \cdot BE_{Dummy, i} + hs \cdot HSE_i + \epsilon_i $.\\ $log(BE/ME)^{+}_i$ is BE/ME value that fills BE/ME with BE less than 0. $BE_{Dummy, i}$ is 1 when BE is less than 0, else 1. Then we calculated time-series average and its t-value of coefficients and intercept. Returns are observed at July, year $t$ to June, year $t+1$, and BE/ME are observed at December, year $t-1$ and HSE, ME is observed at June, year $t$.}
\label{vg}
\end{table}

\section{Conclusion}

Here we conclude the paper by exploring alternative interpretation of the result, and application of the result to investors.

\subsection{What is HSE?}
Here, we need to considerr other interpretation of HSE other than the proxy of demand pressure sensitivity for economic demand. We can hypothesize two theories. First, HSE could be some sort of behavioral error predictor driven by sentiment, such that the predictor predicts the short-term error or long-term market correction dynamics. Second theory is that HSE is really fundamentally related to individual firm, which captures the company's ability to satisfy societies' utility demand. As big firms consist most of the market and economic activities,  the society's happiness demand will be more dependent to to big firms than to small firms. The same logic could be applied to value firms as well.
The first hypothesis is questionable because of many reasons. The most prominent one is that if the theory is correct, the coefficient should be negative, not positive. Even if the behavioral error cannot be adjusted immediately, the phenomena is unlikely to persist for a year to make the coefficient positive.
Big and Small firm Subsample and Value and Growth firm Subsample Fama-Macbeth regression support the latter hypothesis.

Here we can see that there is nonlinar relationship between size or BE/ME and HSE, where HSE explains big and value firm's cross-sectional return difference more better than small and growth firms. This supported by the fact that most of societies utility demand is processed by big firms and mature industries.

\subsection{Application}

As HSE shows explanatory power to big and value firms, this have substantial implication to institutional investors. As institutional investors have out-performance demand and liquidity constraints, they are drawn to big and value firms. Here by applying HSE related strategies, institutional investors can seek for out-performance while maintaining its investment portfolio style. \\
Although because of technical issues, current HSE based on Google Trend data is unstable and have some estimation errors. For applying to real world problems, there should be a robust way of estimating HSE.
In academic perspective, this research has shown that google trend could be more than sentiment proxy. When chosen wisely, some search topics could be related to risk that will be priced. 

\vskip 0.2in

\bibliography{main}
\bibliographystyle{plain}
\end{document}